\documentstyle[prd,preprint,aps,epsfig]{revtex}
\begin{document}
\def\be{\begin{equation}}
\def\ee{\end{equation}}
\def\bea{\begin{eqnarray}}
\def\eea{\end{eqnarray}}
\def\nn{\nonumber}
\def\ep{\epsilon}
\def\ga{\gamma}
\def\Ga{\Gamma}
\def\la{\lambda}
\def\si{\sigma}
\def\al{\alpha}
\def\pa{\partial}
\def\de{\delta}
\def\De{\Delta}
\def\rsr{{r_{s}\over r}}
\def\rrs{{r\over r_{s}}}   
\def\rs2r{{r_{s}\over 2r}}
\def\l2r2{{l^{2}\over r^{2}}}
\def\rsa{{r_{s}\over a}}
\def\rsb{{r_{s}\over b}}
\def\rsro{{r_{s}\over r_{o}}}
\def\rss{r_{s}}
\def\a2{{l^{2}\over a^{2}}}
\def\b2{{l^{2}\over b^{2}}}
\def\op{\oplus}
\def\rab{_{{A}}^{{B}}}

\title{Some Remarks on the Neutrino Oscillation \\
Phase  in a Gravitational Field}

\author{J. G. Pereira and C. M. Zhang}
\vskip 0.5cm
\address{Instituto de F\'{\i}sica Te\'orica\\
Universidade Estadual Paulista\\
Rua Pamplona 145\\
01405-900\, S\~ao Paulo \\
Brazil}
\maketitle

\begin{abstract}
The weak gravitational field expansion method to account for the
gravitationally induced neutrino oscillation effect is critically
examined. It is shown that the splitting of the neutrino phase
into a ``kinematic'' and a ``gravitational'' phase is not always possible
because the relativistic factor modifies the particle interference phase
splitting condition in a gravitational field. 

\end{abstract}
\vskip 1.5cm

\hskip 1,2cm Key words: neutrino oscillation, interference phase, weak field.

\newpage
 
The gravitationally induced neutrino oscillation phase  has
attracted  much attention in recent years
~\cite{ahl96,ahl97,ahl98,ful96,for96,bha99,gro96,wud91,wud96,bru98,koj96}.
However, a lot of problems on the weak field expansion method still exists.  
In order to clarify some of the conceptual problems appearing on the interplay
of gravity and neutrino oscillation, 
we critically examine in this paper the weak-field expansion method 
used to take into account the gravitational effect on the neutrino
oscillation phase. 
We set $G = \hbar = c = 1$ throughout the manuscript.

For one massive neutrino produced at the source position $A$, with  the
detector at position $B$, the 
geometrical optics phase in a curved spacetime can be
calculated by the conventional formula~\cite{ful96,wud91,sto79,ana,aud81}
\be\label{phi}
\Phi = \int\rab m ds = \int\rab
g_{\mu\nu}P^{\mu}dx^{\nu} \; ,
\ee
where $P^{\nu}$ is the 4-momentum, $g_{\mu\nu}$ is the metric
and $ds$ is the spacetime line element. For the case of
two massive neutrinos interference problem, however, a covariant neutrino
wave-packet approach should be introduced to calculate $\Phi$. 

To study the thermal neutron interference of COW experiment~\cite{cow}
in the weak field limit~\cite{sto79}, the phase factor (\ref{phi}) can
still be used. In the Earth's gravitational field, the neutron interference
phenomenon is usually calculated by inserting a Newtonian
potential in the Schroedinger equation~\cite{sak}. This, however,
precludes the description of such interference effect when the gravitational
field is not Newtonian. Another approach to
calculate the thermal neutron interference phase, 
which takes into account the full tensor character of gravitation, can
alternatively be used~\cite{sto79}. 
Following this approach, we consider the linearized gravitational
field (weak field), and  call $h_{\mu\nu}$ the small deviation from the
Minkowski metric $\eta_{\mu \nu}$, so that 
$g_{\mu \nu}= \eta_{\mu \nu} + h_{\mu\nu}$, where 
$h_{\mu\nu} = - \alpha \delta_{\mu\nu}$, with 
$\alpha = {r_{s} / r}$ and 
$r_{s} = 2 M$ the Schwarzschild radius of the
gravitational source mass $M$. When dealing with the thermal neutron
interference, the phase of Eq.(\ref{phi}) may be split up into a
``kinematic'' phase
$\Phi^{0}$ and an extra ``gravitationally induced'' phase 
$\Phi^{G}$~\cite{sto79}:
\be 
\Phi = \Phi^{o} + \Phi^{G} \; .
\ee 
Writing 
\be \label{dsdso}
ds^{2} = (ds^{o})^{2} + h_{\mu\nu} dx^{\mu} dx^{\nu} \; , 
\ee
where $(ds^{o})^{2} = \eta_{\mu\nu} dx^{\mu} dx^{\nu}$ is the flat spacetime
interval, the weak field induced interference phase splitting expansion
reads
\be
ds =[(ds^{o})^{2} + h_{\mu \nu} dx^{\mu} dx^{\nu}]^{1/2}
    \approx ds^{o} + {1 \over 2}h_{\mu \nu} {dx^{\mu}\over ds^o}
    dx^{\nu} \; . \label{dsdso2}
\ee
We then find that $\Phi^{o} = \int\rab m ds^{o}$ is the phase in the flat
spacetime, and 
\be 
\Phi^{G} = {1 \over 2} \int\rab h_{\mu\nu} P_{(o)}^{\mu} dx^{\nu} \; ,
\ee
where $P_{(o)}^{\mu}= m dx^{\mu}/ds^{o}$, is the usual 4-momentum of special
relativity. For the case of neutrons with not too large translational
velocities ($v^{2} \sim 10^{-10}$)~\cite{sto79,cow}, the above treatment
is
completely equivalent to that using the Newtonian potential, and accounts
correctly for the thermal neutron interference experiment~\cite{sto79}.  

The weak field approximation is a powerful tool to work out gravitationally
related problems~\cite{mtw}, especially those related to neutron
optics~\cite{sto79,ana,cow}. On the other hand, when dealing with the
neutrino phase factor, the weak field approximation 
to account for the interference phase splitting, given by Eqs.(\ref{dsdso}) and
(\ref{dsdso2}), should be carefully considered. Before getting to the point, it
is helpful to keep in mind that, unlike the thermal neutron, massive neutrinos 
propagate at nearly the speed of light, being a
relativistic object with much higher momentum-energy than its rest
mass-energy. 

The weak field induced phase splitting  originates from the fact that the
first term in the
r.h.s. of Eq.(\ref{dsdso}) is much greater than the second one. This
approximation is satisfactorily  applied to the case of the  neutron
optics because thermal neutrons are
low-energy objects. But, for an extremely relativistic object, the weak
field induced phase splitting requires further considerations. In order to
compare the magnitudes
of the two terms in the r.h.s. of Eq.(\ref{dsdso}), we take the ratio
between them 
\be \label{ratio} 
\xi = {|h_{\mu \nu} dx^{\mu} dx^{\nu}| \over (ds^{o})^{2}}
= {\alpha \delta_{\mu\nu} dx^{\mu} dx^{\nu} \over  (ds^{o})^{2}}
= \frac{\alpha \delta_{\mu\nu} dx^{\mu}
dx^{\nu}}{ds^{2}}\frac{ds^2}{(ds^{o})^{2}} .
\ee
We can then write
\be\label{dsdso3}
\frac{ds}{ds^{o}} = \sqrt{1 - \xi} .
\ee
and consequently, we obtain
\be
\xi = \frac{\alpha}{m^2} \, (1 - \xi) \, [(P^{o})^{2} + (P^{r})^{2}] ,
\ee
where the momentum is defined as $P^{\mu} = m (dx^{\mu} / ds)$, 
and $(P^{r})^{2}= (P^{x})^{2}+(P^{y})^{2}+(P^{z})^{2}$. We then get
\be \label{xi}
\xi =\frac{\alpha (2\gamma^{2} - 1)}{1 + \alpha(2\gamma^{2} - 1)} \; ,
\ee
where $\gamma =(P^{o} / m)$ is the relativistic factor. The
approximated  mass shell condition 
$(P^{o})^{2} - (P^{r})^{2} \approx m^{2}$ has been used in the above
expression. 

From Eq.(\ref{dsdso3}) we can see that the value of $\xi$ must be in the
range $0 \le \xi \le 1$, where the values 0 and 1 represent respectively
the vacuum case and the null case (light trajectory). 
In the absence of gravitational field, which corresponds to $\alpha
\rightarrow 0$, we find $\xi = 0$, and the vacuum situation is recovered. 
When $v \rightarrow c $, which corresponds to an ultra high-energy case,
$\gamma \rightarrow \infty$, and we obtain $\xi = 1$. So, we see from
Eq.(\ref{xi}) that in fact $0 \le \xi \le 1$ in any situation. 

We can now discuss the expansion conditions. For a low energy object, as
for example a thermal neutron in the laboratory, whose typical velocity
is $v^2 \sim 10^{-10}$, we have $\gamma \simeq 1$, and consequently
\be
\xi =\frac{\alpha }{1 + \alpha} \simeq  {r_{s}\over r} \ll 1 .
\ee
This is the conventional weak field condition, 
which means that the phase splitting can be performed. 
On the other hand, for high-energy massive neutrinos, as
for example an electron neutrino, the relativistic factor is 
$\gamma^{2} \sim 10^{12}$~\cite{boe92,bah94}, and in the case of the
Earth gravitational potential, for which $(r_{s} / r) \sim 10^{-11}$, 
we get $\xi = 0.95 \approx 1$. Consequently, the phase splitting
of Eq.(\ref{dsdso2}) or Eq.(\ref{dsdso3}) can not be performed.  For a
galaxy, the sun, a white dwarf and
a neutron star, the gravitational potentials are respectively $10^{-7}$,
$10^{-6}$, $10^{-3}$ and $10^{-1}$. In all these cases, the electron neutrino
does not admit the conventional phase splitting as the thermal neutron
does. We can then conclude that the particle interference phase splitting
depends on the relativistic factor, and that any interference phase
splitting for high energy particles which does not take into consideration
the relativistic factor, will be meaningless.

Neutrinos are extremely relativistic particles
with very large relativistic factors. Consequently, for these particles, the
interference phase splitting will not work for most of the usual
astrophysical situations. In other words, the conventional interference phase
splitting condition applied to
the thermal neutron interference experiment can not be directly applied to a
neutrino because the relativistic factor modifies the
interference phase splitting  condition. As an immediate consequence, the
splitting of
the neutrino phase into ``kinematic" and ``gravitational" phases is not 
possible for both the neutron star and the Earth gravitational fields,
as is sometimes claimed in the literature
~\cite{ahl96,ahl97,ahl98}. 
On the other hand,
thermal neutrons in COW experiment are low energy objects whose
relativistic factors are nearly unity, 
and the gravity weak-field limit  
yields a critical gravitational potential of
order $(r_{s} / r)_{critical} \sim 1$. For these particles, therefore, one
does not need to consider the relativistic factor in the 
interference phase splitting.

As a further remark, we note that the phase factors $\Phi$ of
Eq.(\ref{phi}) is not the total phase of the neutrino in a gravitational
field, but only the phase due to the coupling of the energy-momentum of
the particle to the (curved) spacetime  geometry, and named type-I
phase~\cite{ana}. Concerning the spinning aspect of the particle, 
gravitation gives rise also to a type-II phase, a phase shift related to
the coupling of the ``spin connection" to the curvature. The general physical
characters of these two types of phase factors can be found in
Ref.~\cite{ana}. It should be mentioned, however, that the
type-II phase has no contribution to the neutrino oscillation phase in a
Schwarzschild spacetime because of the static and spherical
symmetry~\cite{ful96}. 

\section*{Acknowledgments}

One of the authors (JGP) would like to thank CNPq--Brazil for partial
financial support. The other (CMZ) would like to thank FAPESP-Brazil for
financial support. They would like also to thank M. Nowakowski for 
helpful discussions, and G. F. Rubilar for valuable comments.

\end{document}